\documentclass[conference]{IEEEtran}
\IEEEoverridecommandlockouts
\usepackage{cite}
\usepackage{amsmath,amssymb,amsfonts}
\usepackage{graphicx}
\usepackage{textcomp}
\usepackage{xcolor}

\usepackage{dcolumn}
\usepackage{bm}
\usepackage{qcircuit}
\usepackage{braket}
\usepackage{float}
\usepackage{algorithm}
\usepackage{algorithmicx}
\usepackage{algpseudocode}
\usepackage{mathtools}
\usepackage{hyperref}
\usepackage{multirow}

\renewcommand{\figureautorefname}{Figure~\negthinspace}

\renewcommand{\tableautorefname}{Table~\negthinspace}
\renewcommand{\sectionautorefname}{Section~\negthinspace}

\def\BibTeX{{\rm B\kern-.05em{\sc i\kern-.025em b}\kern-.08em
    T\kern-.1667em\lower.7ex\hbox{E}\kern-.125emX}}
\begin{document}

\title{Learning to Program Variational Quantum Circuits with Fast Weights

\thanks{The views expressed in this article are those of the authors and do not represent the views of Wells Fargo. This article is for informational purposes only. Nothing contained in this article should be construed as investment advice. Wells Fargo makes no express or implied warranties and expressly disclaims all legal, tax, and accounting implications related to this article.}
}

\author{\IEEEauthorblockN{ Samuel Yen-Chi Chen}
\IEEEauthorblockA{
\textit{Wells Fargo}\\
New York, NY, USA \\
yen-chi.chen@wellsfargo.com}
}


\maketitle

\begin{abstract}

Quantum Machine Learning (QML) has surfaced as a pioneering framework addressing sequential control tasks and time-series modeling. It has demonstrated empirical quantum advantages notably within domains such as Reinforcement Learning (RL) and time-series prediction. A significant advancement lies in Quantum Recurrent Neural Networks (QRNNs), specifically tailored for memory-intensive tasks encompassing partially observable environments and non-linear time-series prediction. Nevertheless, QRNN-based models encounter challenges, notably prolonged training duration stemming from the necessity to compute quantum gradients using backpropagation-through-time (BPTT). This predicament exacerbates when executing the complete model on quantum devices, primarily due to the substantial demand for circuit evaluation arising from the parameter-shift rule. This paper introduces the Quantum Fast Weight Programmers (QFWP) as a solution to the temporal or sequential learning challenge. The QFWP leverages a classical neural network (referred to as the 'slow programmer') functioning as a quantum programmer to swiftly modify the parameters of a variational quantum circuit (termed the 'fast programmer'). Instead of completely overwriting the fast programmer at each time-step, the slow programmer generates parameter changes or updates for the quantum circuit parameters. This approach enables the fast programmer to incorporate past observations or information. Notably, the proposed QFWP model achieves learning of temporal dependencies without necessitating the use of quantum recurrent neural networks. Numerical simulations conducted in this study showcase the efficacy of the proposed QFWP model in both time-series prediction and RL tasks. The model exhibits performance levels either comparable to or surpassing those achieved by QLSTM-based models.

\end{abstract}


\section{\label{sec:Indroduction}Introduction}
\begin{figure}[htbp]
\begin{center}
\includegraphics[width=1\columnwidth]{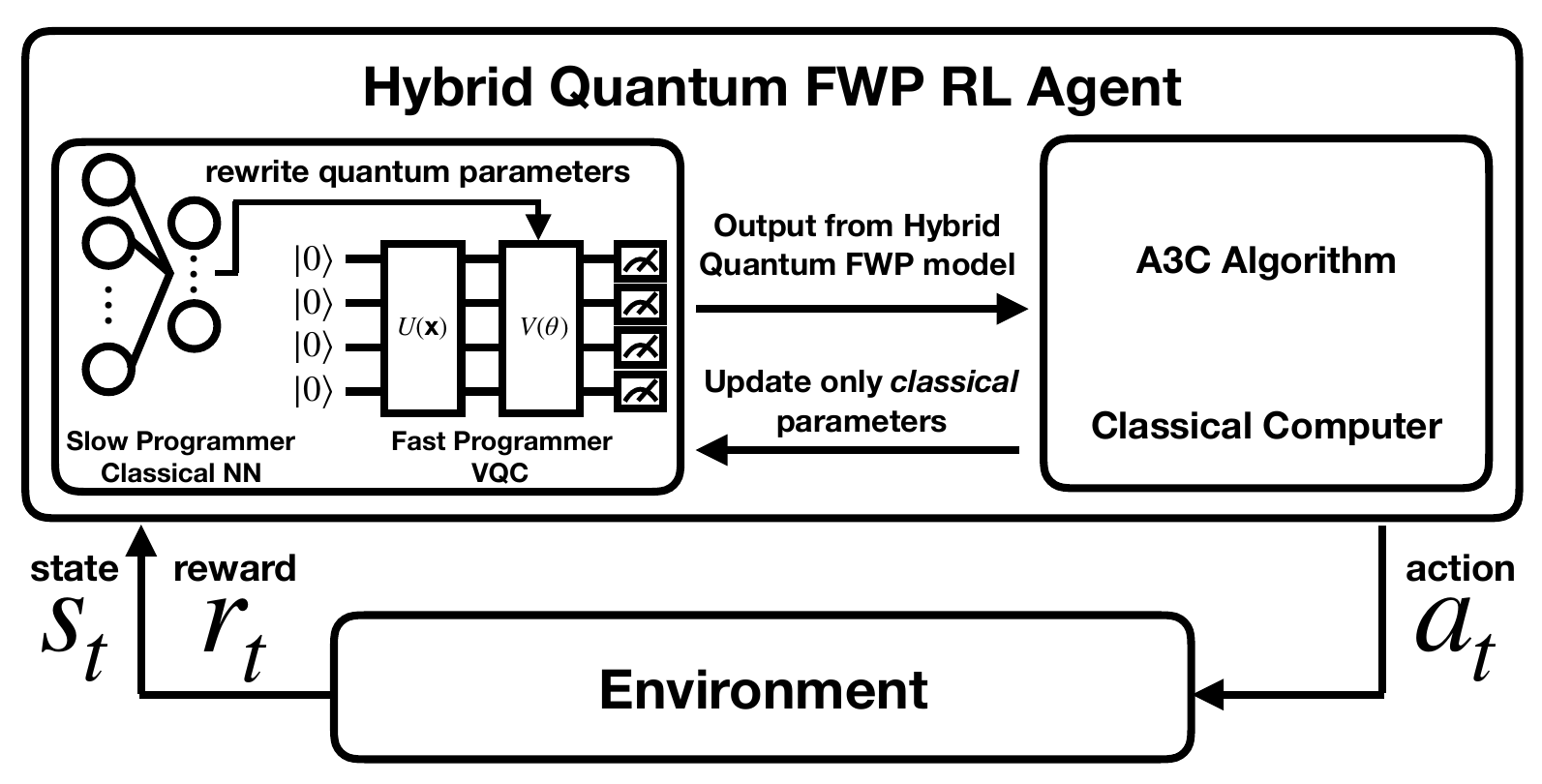}\vskip -0.1in
\caption{{\bfseries Hybrid Quantum Fast Weight Programmer (FWP) as a RL agent.}}
\label{fig:generic_FWP_RL}
\end{center}
\end{figure}

While quantum computing (QC) holds promise for excelling in complex computational tasks compared to classical systems \cite{nielsen2010quantum}, the current challenge lies in the absence of error correction and fault tolerance, complicating the implementation of deep quantum circuits for complex quantum algoruthms. These noisy intermediate-scale quantum (NISQ) devices \cite{preskill2018quantum} require specialized quantum circuit designs to fully harness their potential advantages. A recent hybrid quantum-classical computing approach \cite{bharti2022noisy} strategically combines both realms: quantum computers handle advantageous tasks, while classical counterparts manage computations, such as gradient calculations. Known as \emph{variational quantum algorithms} (VQA), these methods have excelled in various machine learning (ML) tasks such as classification \cite{mitarai2018quantum,qi2023qtnvqc,chen2021end}, sequential modeling/control \cite{chen2022quantumLSTM,bausch2020recurrent}, generative models \cite{chu2023iqgan,stein2021qugan} and natural language processing \cite{yang2021decentralizing,li2023pqlm,yang2022bert,di2022dawn,stein2023applying}. Time-series modeling and reinforcement learning (RL) within the realm of ML demand specific memory mechanisms to retain past observations for optimal performance. Classical ML fields have extensively explored these tasks using deep neural networks, achieving significant progress as evidenced by seminal works \cite{hochreiter1997long,vaswani2017attention,mnih2015human}. However, the exploration of these tasks within QML remains largely uncharted territory. Current QML methods designed for addressing time-series modeling or RL tasks, demanding the retention of past observations, such as those utilizing quantum recurrent neural networks (QRNNs), encounter prolonged training time. This challenge arises from the necessity to compute quantum gradients across extensive or deep circuits and the substantial quantum circuit evaluations required to gather expectation values. This paper presents the \emph{Quantum Fast Weight Programmers} (QFWP), a framework utilizing a classical NN termed the 'slow programmer' to efficiently adjust the parameters of a quantum circuit known as the 'fast programmer'. This methodology is proposed to address challenges in time-series prediction and RL without relying on QRNN, as depicted in \figureautorefname{\ref{fig:generic_FWP_RL}}. Our numerical simulations demonstrate that the proposed QFWP framework achieves performance levels comparable to or surpassing those of fully trained QRNN counterparts, such as QLSTM, under equivalent model sizes and training configurations.
\section{\label{sec:Related_Work}Related Work}
Quantum reinforcement learning (QRL) has been an area of study since the seminal work by Dong et al. in 2008 \cite{dong2008quantum}. Initially, its applicability was constrained by the necessity to construct the environment in a completely quantum manner, limiting its practical utility. Subsequent advancements in QRL, leveraging Variational Quantum Circuits (VQCs), have extended its scope to address classical environments with both discrete \cite{chen19} and continuous observation spaces \cite{lockwood2020reinforcement, skolik2021quantum}. The progression of QRL has seen enhancements in performance through the adoption of policy-based learning approaches, including Proximal Policy Optimization (PPO) \cite{hsiao2022unentangled}, Soft Actor-Critic (SAC) \cite{lan2021variational}, REINFORCE \cite{jerbi2021variational}, Advantage Actor-Critic (A2C) \cite{kolle2024quantum} and Asynchronous Advantage Actor-Critic (A3C) \cite{CHEN2023321Async}. Furthermore, to address the challenges posed by partially observable environments, researchers have employed quantum recurrent neural networks as reinforcement learning policies \cite{chen2023quantum_LSTM_RL,chen2023efficientQRL_QRC}. Time-series modeling and prediction represent significant application scenarios within QML. Recent investigations in QML have drawn inspiration from successful methodologies in classical ML, particularly RNN-based approaches. Quantum RNNs have exhibited promise in accurately modeling time-series data, as evidenced by notable works \cite{bausch2020recurrent,chen2022quantumLSTM, chen2022reservoir}. However, the training of quantum RNNs is susceptible to prolonged duration due to the necessity of backpropagation-through-time (BPTT) \cite{chen2022quantumLSTM} and the potential involvement of deep quantum circuits \cite{bausch2020recurrent}.
The primary aim of the proposed QFWP model is to emulate the memory capabilities inherent in QRNNs while eliminating the requirement for recurrent connections, a notable factor contributing to prolonged training periods as emphasized in \cite{chen2023quantum_LSTM_RL,chen2023efficientQRL_QRC}. Notably, the QFWP model demonstrates comparable time-series modeling proficiency to QRNN-based models without necessitating intricate backpropagation-through-time (BPTT) across quantum circuits or the use of deep quantum circuits. This study employs the A3C RL learning algorithm \cite{mnih2016asynchronous,CHEN2023321Async}, optimizing the utilization of multiple-core CPU computing resources. This choice underscores potential applications in scenarios involving arrays of quantum computers.
The work in \cite{verdon2019learning} introduces a framework employing a classical RNN for optimizing VQC parameters. This method involves utilizing VQC parameters and observable outputs from preceding time-steps as inputs to the RNN, generating subsequent VQC parameters. Such an approach demonstrates proficient parameter initialization for various VQC-based tasks. Our approach differs significantly from the aforementioned method as we do not employ an RNN to generate quantum parameters. Instead, we utilize a simple feed-forward NN solely responsible for updating quantum parameters. Furthermore, our approach does not necessitate feeding quantum parameters from the previous time-step into the NN for generating new parameters. The motivation behind our approach is to simplify the classical NN to minimize computational load while achieving our objectives.

\section{\label{sec:FWP}Fast Weight Programmers}
\begin{figure}[htbp]
\begin{center}
\includegraphics[width=1\columnwidth]{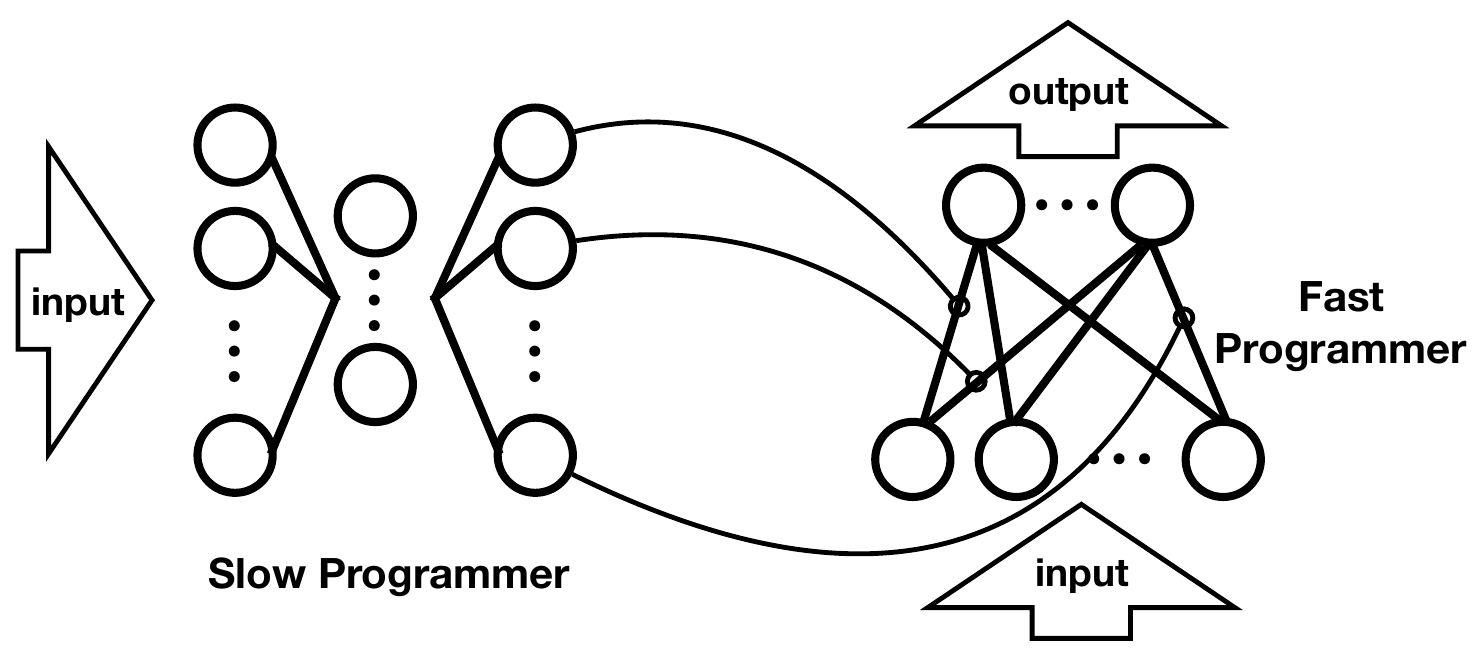}\vskip -0.1in
\caption{{\bfseries Generic Structure of a Fast Weight Programmer (FWP).}}
\label{fig:generic_FWP}
\end{center}
\vskip -0.1in
\end{figure}

The concept of \emph{Fast Weight Programmers} (FWP), illustrated in \figureautorefname{\ref{fig:generic_FWP}}, was initially introduced in the works by Schmidhuber \cite{schmidhuber1992learning,schmidhuber1993reducing}. In this sequential learning model, two distinct neural networks (NN) are employed, termed the \emph{slow programmer} and the \emph{fast programmer}. The NN weights in this context serve as the model/agent's \emph{program}. The fundamental idea of FWP involves the slow programmer, during a given time-step, generating \emph{updates} or \emph{changes} to the NN weights of the fast programmer based on observations. This \emph{reprogramming} process swiftly redirects the fast programmer's focus to more salient information within the incoming data stream. It is crucial to note that the slow programmer does not entirely overwrite the fast programmer; rather, only changes or updates are applied. This method enables the fast programmer to consider previous observations or information, offering a mechanism for a simple feed-forward NN to handle sequential prediction or control without resorting to recurrent neural networks (RNN), which typically demand significant computational resources.
In the original configuration, each connection of the fast programmer is associated with a distinct output unit in the slow programmer. The evolution of fast programmer weights is characterized by the update rule $W^{\text{fast}}(t+1) \leftarrow W^{\text{fast}}(t) + \Delta W(t)$, where $\Delta W(t)$ denotes the output from the slow programmer at time-step $t$.
The original scheme may encounter scalability issues when the fast programmer NN is large, as the slow programmer requires an equal number of output neurons as the connections in the fast programmer NN. An alternative method is proposed in \cite{schmidhuber1992learning}, where the slow NN includes a specialized unit for each unit in the fast NN, labeled as \emph{FROM}, corresponding to units from which at least one fast NN connection originates. Similarly, the slow NN has a special unit for each unit in the fast NN, labeled as \emph{TO}, corresponding to units to which at least one fast NN connection leads. Under this configuration, updates for fast NN weights are computed as $\Delta W_{ij}(t) = W^{\text{FROM}}_{i}(t) \times W^{\text{TO}}_{j}(t)$, where $\Delta W_{ij}(t)$ signifies the update for fast NN weight $W^{\text{fast}}_{ij}(t)$. The entire FWP model can be optimized end-to-end using gradient-based or gradient-free methods. FWP has demonstrated effectiveness in solving time-series modeling \cite{schmidhuber1992learning} and reinforcement learning tasks \cite{gomez2005evolving}. The proposed Quantum Fast Weight Programmer (QFWP) model draws inspiration from the original FWP, with a generalization that the fast programmer can be implemented using trainable quantum circuits, as detailed in \sectionautorefname{\ref{sec:QuantumFWP}}.

\section{\label{sec:VQC}Variational Quantum Circuits}
\begin{figure}[htbp]
\begin{center}
\includegraphics[width=1\columnwidth]{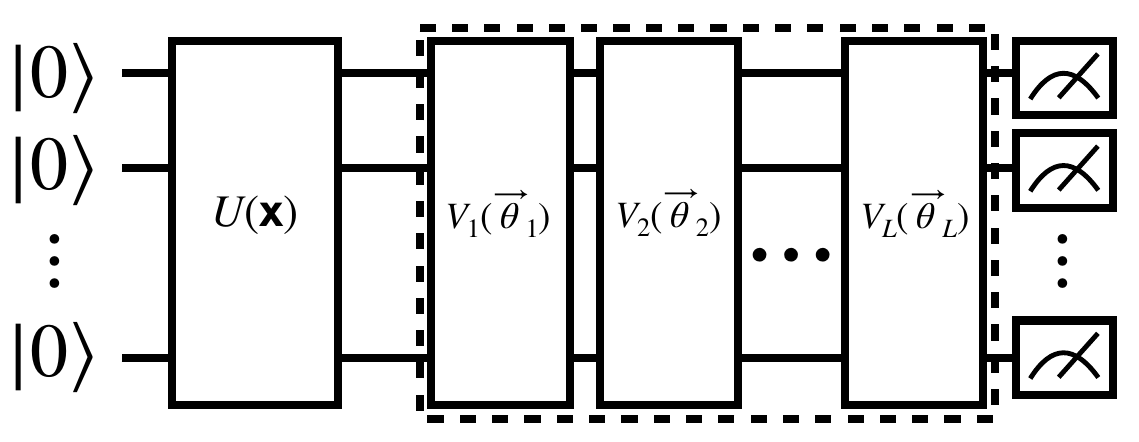}\vskip -0.1in
\caption{{\bfseries Generic Structure of a Variational Quantum Circuit (VQC).}}
\label{fig:generic_VQC}
\end{center}
\vskip -0.1in
\end{figure}

\begin{figure}[htbp]
\begin{center}
\includegraphics[width=1\columnwidth]{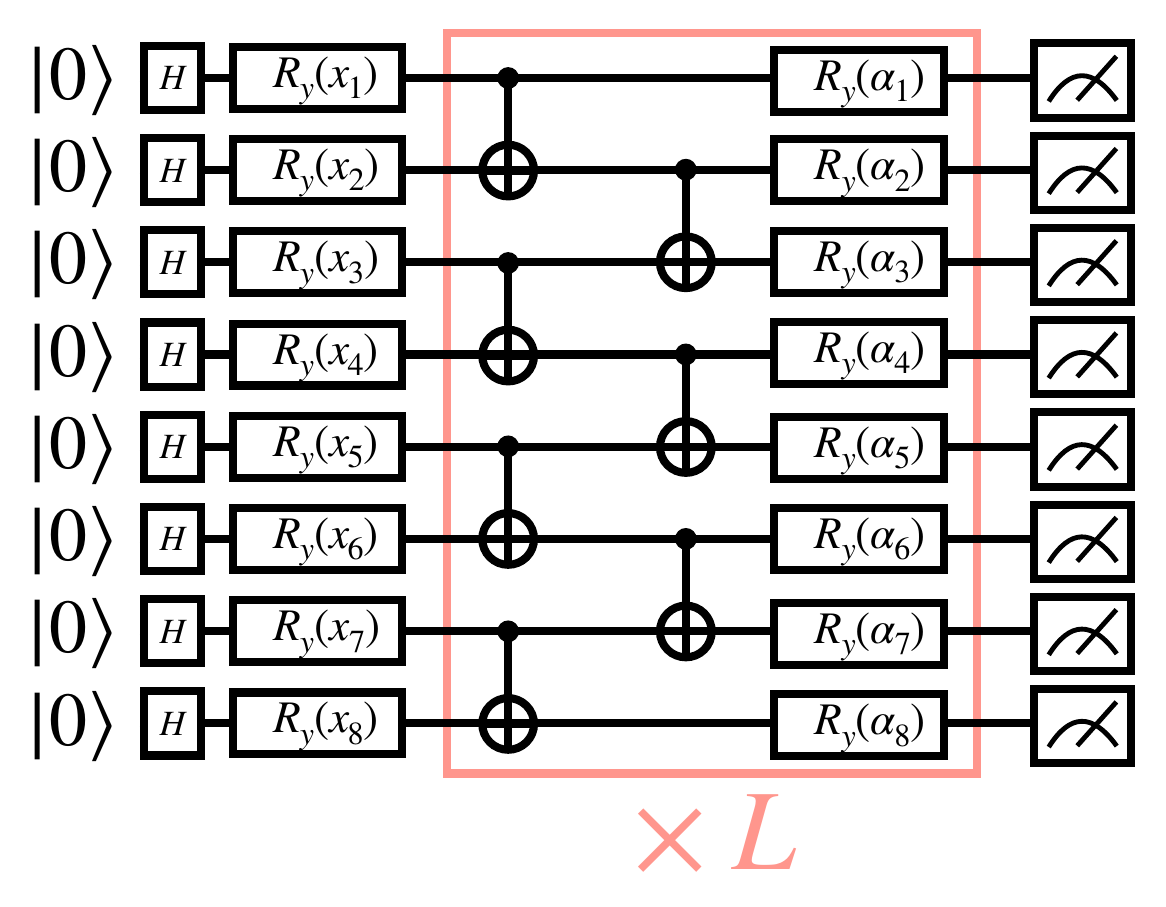}\vskip -0.1in
\caption{{\bfseries VQC used in this paper.}}
\label{fig:detailed_VQC}
\end{center}
\vskip -0.1in
\end{figure}

\emph{Variational quantum circuits} (VQC), also known as \emph{parameterized quantum circuits} (PQC), is a special kind of quantum circuit with trainable parameters. VQC has been used widely in current hybrid quantum-classical computing framework \cite{bharti2022noisy} and shown to have certain kinds of quantum advantages \cite{abbas2021power,caro2022generalization,du2020expressive}. 
In general, a VQC includes three fundamental components: \emph{encoding circuit}, \emph{variational circuit} and the final \emph{measurements}.
As shown in \figureautorefname{\ref{fig:generic_VQC}}, the encoding circuit $U(\mathbf{x})$ transforms the initial quantum state $\ket{0}^{\otimes n}$ into $\ket{\Psi} = U(\mathbf{x})\ket{0}^{\otimes n}$. Here the $n$ represents the number of qubits and the $U(\mathbf{x})$ represents the unitary which depends on the input value $\mathbf{x}$. 
The VQC used in this paper is shown in \figureautorefname{\ref{fig:detailed_VQC}}. The encoding layer $U(\mathbf{x})$ includes Hadamard gates on all qubits to initialize an unbiased state $H\ket{0}\otimes \cdots \otimes  H\ket{0} = \sum_{(q_1,q_2,...,q_n) \in \{ 0,1\}^n} \frac{1}{\sqrt{2^n}} \ket{q_1} \otimes \ket{q_2} \otimes \cdots \otimes \ket{q_n}$ and $R_{y}$ gates with rotation angles corresponding to the input data $x_1, \cdots, x_n$. It can be expressed as $U(\mathbf{x}) = R_{y}(x_1)H \otimes \cdots \otimes R_{y}(x_n)H$.
The encoded state then go through the variational circuit (shown in dashed box) which includes multiple layers of trainable quantum circuits $V_{j}(\Vec{\theta_{j}})$. The $V_{j}$ circuit block used in this work is shown in the boxed region in \figureautorefname{\ref{fig:detailed_VQC}} and can be implemented $L$ repetitions to increase the number of trainable parameters. Each of the $V_{j}$ circuit block includes CNOT gates to entangle quantum information and parameterized $R_{y}$ gates.
The overall action $W(\Theta)$ in the trainable part is therefore $W(\Theta) = V_{L}(\Vec{\theta_{L}})V_{L-1}(\Vec{\theta_{L-1}}) \cdots V_{1}(\Vec{\theta_{1}})$ where $L$ represents the number of layers and $\Theta$ is the collection of all trainable parameters $\Vec{\theta_{1}} \cdots \Vec{\theta_{L}}$. 
The measurement process furnishes data by evaluating a subset or entirety of the qubits, generating a classical bit sequence for subsequent utilization. Executing the circuit once produces a bit sequence such as "0,0,1,1." Yet, conducting the circuit multiple times (shots) provides expectation values for each qubit. This study particularly focuses on the evaluation of Pauli-$Z$ expectation values derived from measurements in VQCs.
The mathematical expression of the VQC used in this work is $\overrightarrow{f(\mathbf{x} ; \Theta)}=\left(\left\langle\hat{Z}_1\right\rangle, \cdots,\left\langle\hat{Z}_n\right\rangle\right)$ , where $\left\langle\hat{Z}_{k}\right\rangle =\left\langle 0\left|U^{\dagger}(\mathbf{x})W^{\dagger}(\Theta) \hat{Z_{k}} W(\Theta)U(\mathbf{x})\right| 0\right\rangle$.

In the hybrid quantum-classical framework, the VQC can be integrated with other classical (e.g. deep neural networks, tensor networks) or quantum  (e.g. other VQCs) components and the whole model can be optimized in an end-to-end manner via gradient-based \cite{chen2021end,qi2023qtnvqc} or gradient-free \cite{chen2022variationalQRL} methods. When gradient-based methods such as gradient descent are used, the gradients of quantum components can be calculated through the parameter-shift rules \cite{mitarai2018quantum,schuld2019evaluating,bergholm2018pennylane}.
In ordinary VQC-based ML models, the parameters $\Theta$ are randomly initialized and then updated iteratively. In our proposed QFWP model, these quantum circuit parameters, or more specifically, the \emph{updates} or \emph{changes} of the quantum circuit parameters, are the output from another classical NN (\emph{slow programmer} in the FWP setting).
\section{\label{sec:Methods}Methods}
\subsection{\label{sec:QuantumFWP}Quantum FWP}
\begin{figure}[htbp]
\begin{center}
\includegraphics[width=1\columnwidth]{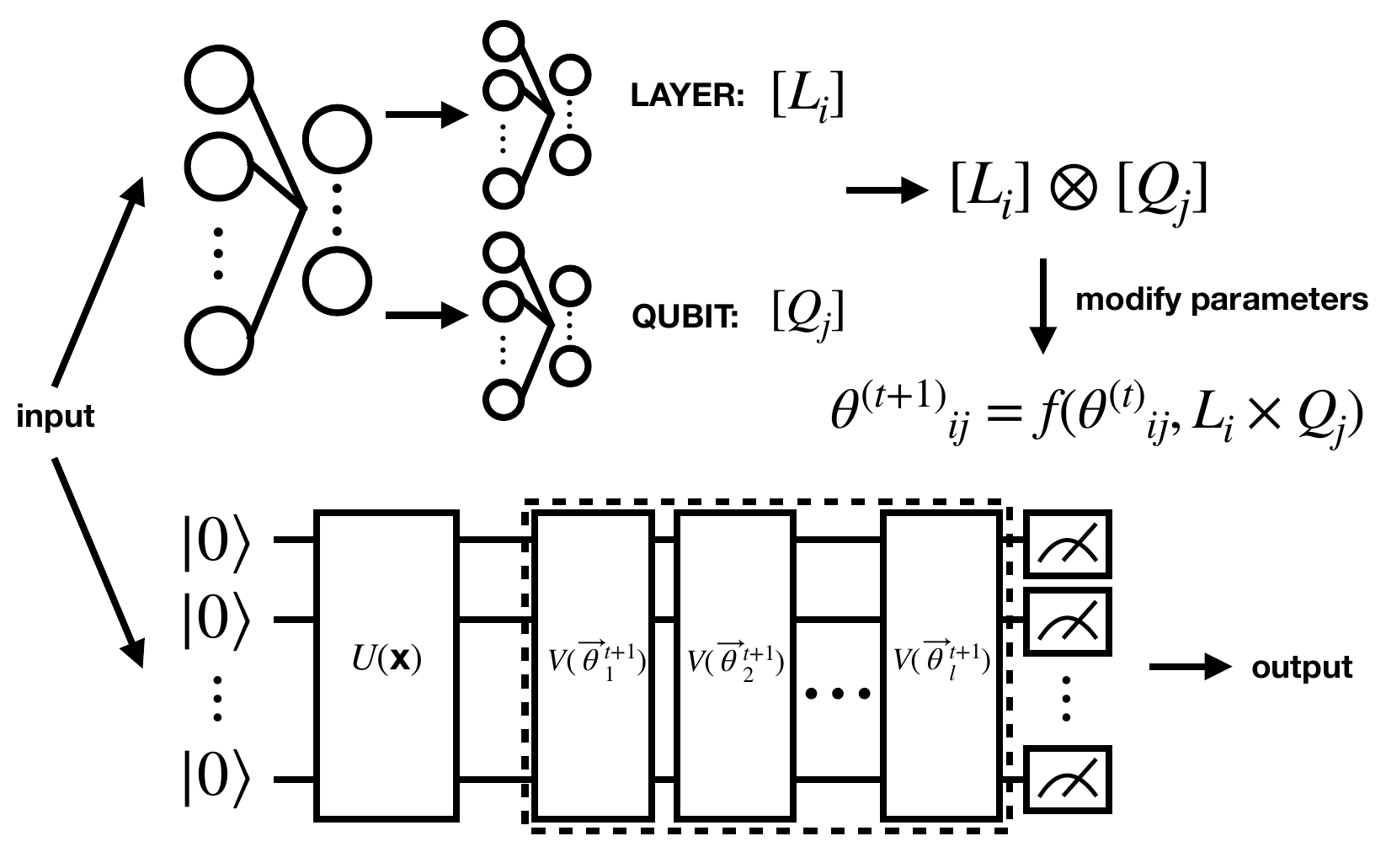}\vskip -0.1in
\caption{{\bfseries Quantum Fast Weight Programmers}}
\label{fig:vqFWP_Concept}
\end{center}
\vskip -0.1in
\end{figure}

In the proposed quantum fast weight programmers (QFWP), we employ the hybrid quantum-classical architecture to leverage the best part from both the quantum and classical worlds. We employ the classical neural networks to build the \emph{slow} networks, which will generate the values to update the parameters of the \emph{fast} networks, which is actually a VQC. As shown in \figureautorefname{\ref{fig:vqFWP_Concept}}, the input vector $\Vec{x}$ is loaded into the classical neural network encoder, the output from the encoder network is then processed by another two neural networks. One of the network will generate an output vector $[L_{i}]$ with the length equals to the number of VQC layers, and the other network will generate an output vector $[Q_{j}]$ with the length equals to the number of qubits of the VQC. We then calculate the outer product of $[L_{i}]$ and $[Q_{j}]$. It can be written as $[L_{i}] \otimes [Q_{j}] = [M_{ij}] = [L_{i} \times Q_{j}] = 
\begin{bmatrix}
L_{1} \times Q_{1} & L_{1} \times Q_{2} & \cdots & L_{1} \times Q_{n}\\
L_{2} \times Q_{1} & L_{2} \times Q_{2} & \cdots & L_{2} \times Q_{n}\\
\vdots             &       \ddots       &        &        \vdots     \\
L_{l} \times Q_{1} & L_{l} \times Q_{2} & \cdots & L_{l} \times Q_{n}\\
\end{bmatrix}$, where $l$ is the number of learnable layers in VQC and $n$ is the number of qubits.
At time $t+1$, we can calculate the updated VQC parameters as $\theta^{t+1}_{ij} = f(\theta^{t}_{ij}, L_{i} \times Q_{j})$, where $f$ is a function to combine the parameters from the previous time-step $\theta^{t}_{ij}$ and the newly computed $L_{i} \times Q_{j}$. In the time series modeling and RL tasks considered in this work, we adopted the \emph{additive} update rule in which the new circuit parameters are calculated according to $\theta^{t+1}_{ij} = \theta^{t}_{ij} + L_{i} \times Q_{j}$. Through this method, the information from previous time steps can be kept in the form of circuit parameters and affect the VQC behavior when a new input $\Vec{x}$ is given.
The output from the VQC can be further processed by other components such as scaling, translation or a classical neural network to refine the final results.

\section{\label{sec:Experiments}Experiments}
In this research, we use the following open-source package to perform the simulation. We use the PennyLane~\cite{bergholm2018pennylane} for the construction of quantum circuits and PyTorch for the overall hybrid quantum-classical model building.

\subsection{Time-Series Modeling}
We explore four distinct cases (Damped SHM, Bessel function, NARMA5, and NARMA10), previously examined in related studies, to assess the performance of the proposed QFWP against quantum RNN-based methods \cite{chen2022quantumLSTM,chen2022reservoir}. The training and testing methodology follows the approach outlined in \cite{chen2022quantumLSTM,chen2022reservoir}. In essence, the model is tasked with predicting the ($N+1$)-th value, given the first $N$ values in the sequence. For instance, if at time-step $t$, the input to the model is $[x_{t-4}, x_{t-3}, x_{t-2}, x_{t-1}]$ (where $N = 4$), the model is expected to generate the output $y_t$, which ideally should closely align with the ground truth $x_{t}$. In all experiments regarding time-series modeling, we set $N=4$. The presented results showcase the ground truth through an orange dashed line, while the blue solid line represents the output from the models. The vertical red dashed line delineates the \emph{training} set (left) from the \emph{testing} set (right). Across all datasets considered in this paper, $67\%$ of the data is allocated for training, with the remaining $33\%$ dedicated to testing. 
The slow programmer NN for time-series modeling tasks within the QFWP comprises an encoder with $1 \times 8 + 8 = 16$ parameters, a NN for layer index with $8 \times 2 + 2 = 18$ parameters, and a NN for qubit index with $8 \times 8 + 8 = 72$ parameters. Considering an 8-qubit VQC with 2 variational layers as the fast programmer, the total quantum parameters subject to slow programmer updating amount to $8 \times 2 = 16$. Measurement is performed on only 4 qubits from the fast programmer, followed by a post-processing NN (with $4 \times 1 + 1 = 5$ parameters) to generate the final result. This post-processing NN follows the same structure as the one employed in \cite{chen2022reservoir}. In the time-series modeling experiments conducted, solely the first qubit is employed for loading the time-series data, given that only one value is present at each time-step. The number of parameters for both the proposed QFWP and the QLSTM model baseline, as reported in \cite{chen2022reservoir}, are detailed in \tableautorefname{\ref{tab:time_series_number_params}}.
\begin{table}[htbp]
\caption{\bfseries{Number of parameters in QFWP and QLSTM models.}}
\begin{tabular}{|l|ll|ll|}
\hline
\multirow{2}{*}{} & \multicolumn{2}{l|}{QLSTM \cite{chen2022reservoir}}               & \multicolumn{2}{l|}{QFWP}                \\ \cline{2-5} 
                  & \multicolumn{1}{c|}{Classical} & Quantum & \multicolumn{1}{c|}{Classical} & Quantum \\ \hline
Damped SHM/Bessel & \multicolumn{1}{c|}{5}&\multicolumn{1}{c|}{144}& \multicolumn{1}{c|}{111}&\multicolumn{1}{c|}{16}\\ \hline
NARMA5/NARMA10    & \multicolumn{1}{c|}{5}&\multicolumn{1}{c|}{288}& \multicolumn{1}{c|}{111}&\multicolumn{1}{c|}{16}\\ \hline
\end{tabular}
\label{tab:time_series_number_params}
\end{table}
\subsubsection{\label{sec:function_approximation_damped_SHM}Function Approximation-Damped SHM}
\begin{figure}[htbp]
\begin{center}
\includegraphics[width=1\columnwidth]{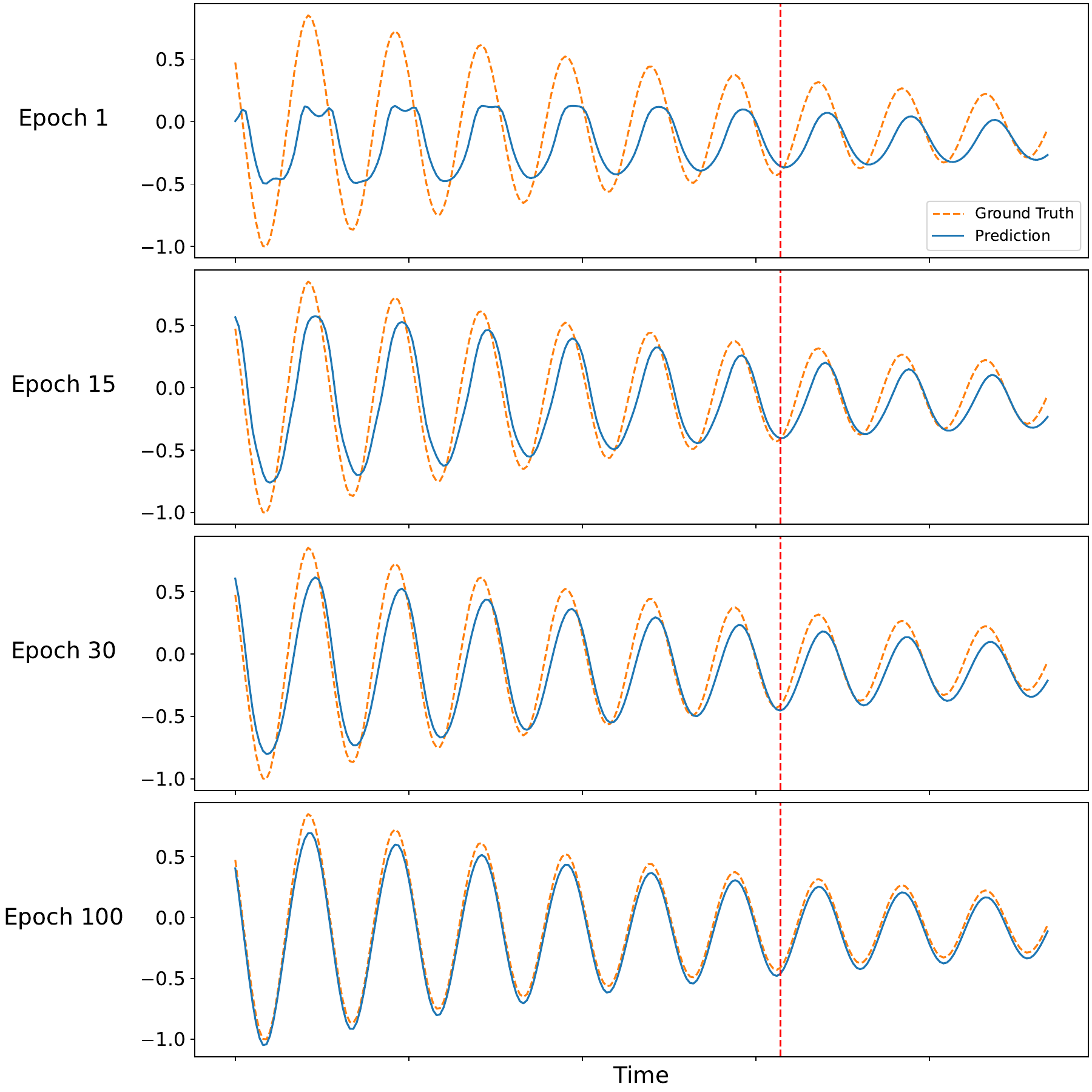}\vskip -0.1in
\caption{{\bfseries Results: Quantum FWP for damped SHM.}}
\label{fig:results_dampedSHM}
\end{center}
\vskip -0.1in
\end{figure}

Damped harmonic oscillators find utility in representing or approximating diverse systems, such as mass-spring systems and acoustic systems. The dynamics of damped harmonic oscillations are encapsulated by the following equation:
\begin{equation}
    \frac{\mathrm{d}^{2} x}{\mathrm{d} t^{2}}+2 \zeta \omega_{0} \frac{\mathrm{d} x}{\mathrm{d} t}+\omega_{0}^{2} x=0,
\end{equation}
where $\omega_{0}=\sqrt{\frac{k}{m}}$ is the (undamped) system's characteristic frequency and $\zeta=\frac{c}{2 \sqrt{m k}}$ is the damping ratio. In this paper, we consider a specific example from the simple pendulum with the following formulation:
\begin{equation}
\frac{d^{2} \theta}{d t^{2}}+\frac{b}{m} \frac{d \theta}{d t}+\frac{g}{L} \sin \theta=0,
\end{equation}
in which the gravitational constant $g = 9.81$, the damping factor $b = 0.15$, the pendulum length $l = 1$ and mass $m = 1$. The initial condition at $t = 0$ has angular displacement $\theta = 0$, and the angular velocity $\dot{\theta} = 3$ rad/sec. We present the QFWP learning result of the angular velocity $\dot{\theta}$.
As depicted in \figureautorefname{\ref{fig:results_dampedSHM}}, the QFWP successfully captures the periodic features after just a single epoch of training, and it approximately captures the essential amplitude features after 15 epochs of training. The performance of QFWP in this context is comparable to the previously reported results of a fully trained QLSTM model in \cite{chen2022reservoir}. While QFWP does not surpass the fully trained QLSTM in performance, it is noteworthy that the function-fitting performance is closely matched, and QFWP is, in fact, a smaller model, as indicated in \tableautorefname{\ref{tab:time_series_number_params}}.
\begin{table}[htbp]
\caption{\bfseries{Results: Time-Series Modeling - Damped SHM}}
\begin{tabular}{|l|l|l|}
\hline
          & QLSTM \cite{chen2022reservoir} & QFWP \\ \hline
Epoch 1   &$1.66 \times 10^{-1}$/$1.35 \times 10^{-2}$&$3.33 \times 10^{-1}$/$3.26 \times 10^{-2}$\\ \hline
Epoch 15  &$2.89 \times 10^{-2}$/$5.53 \times 10^{-3}$&$7.21 \times 10^{-2}$/$1.65 \times 10^{-2}$\\ \hline
Epoch 30  &$9.06 \times 10^{-3}$/$3.41 \times 10^{-4}$&$5.96 \times 10^{-2}$/$1.34 \times 10^{-2}$\\ \hline
Epoch 100 &$2.86 \times 10^{-3}$/$1.94 \times 10^{-4}$&$1.09 \times 10^{-2}$/$2.70 \times 10^{-3}$\\ \hline
\end{tabular}
\label{tab:results_damped_SHM_compare}
\end{table}
\subsubsection{\label{sec:function_approximation_bessel}Function Approximation-Bessel Function}
\begin{figure}[htbp]
\begin{center}
\includegraphics[width=1\columnwidth]{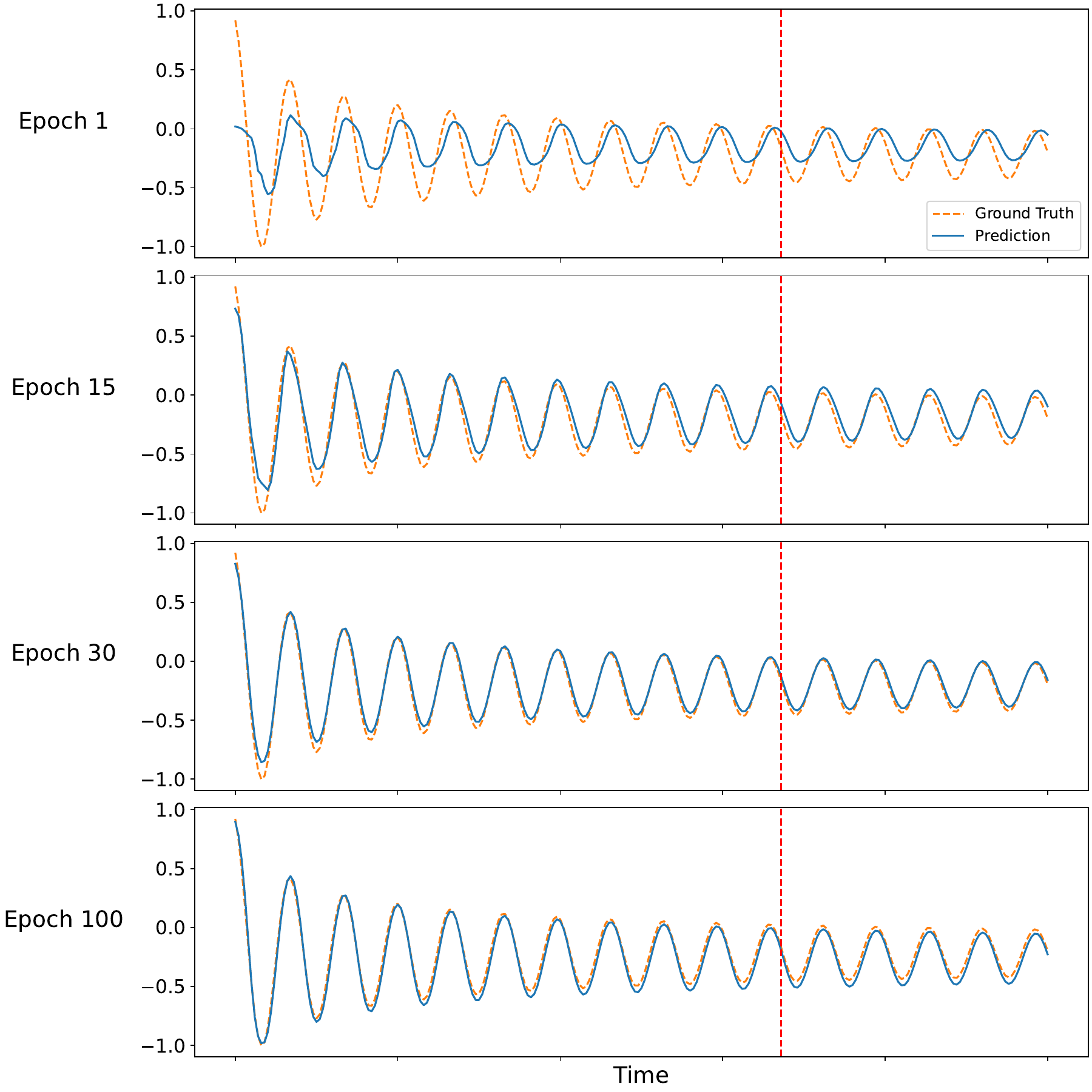}\vskip -0.1in
\caption{{\bfseries Results: Quantum FWP for Bessel function $J_{2}$.}}
\label{fig:results_Bessel}
\end{center}
\vskip -0.1in
\end{figure}

Bessel functions are commonly encountered in various physics and engineering applications, particularly in scenarios such as electromagnetic fields or heat conduction within cylindrical geometries. Bessel functions of the first kind, denoted as $J_\alpha(x)$, serve as solutions to the Bessel differential equation given by:
\begin{equation}
    x^{2} \frac{d^{2} y}{d x^{2}}+x \frac{d y}{d x}+\left(x^{2}-\alpha^{2}\right) y=0,
\end{equation}
and can be defined as 
\begin{equation}
    J_{\alpha}(x)=\sum_{m=0}^{\infty} \frac{(-1)^{m}}{m!  \Gamma(m+\alpha+1)}\left(\frac{x}{2}\right)^{2 m+\alpha},
\end{equation}
where $\Gamma(x)$ is the Gamma function. In this study, we specifically opt for $J_2$ as the function used for QFWP training.
As illustrated in \figureautorefname{\ref{fig:results_Bessel}}, the QFWP adeptly captures periodic features after just a single epoch of training and approximately captures essential amplitude features after 15 epochs. The QFWP demonstrates a nearly perfect approximation of the $J_{2}$ function after 100 epochs of training. As indicated in \tableautorefname{\ref{tab:results_Bessel_compare}}, despite the smaller size of the QFWP model compared to QLSTM, as reported in \cite{chen2022reservoir}, the QFWP achieves performance closely aligned with the previously reported QLSTM results. Notably, at Epochs 15 and 30, the QFWP model achieves training and testing losses lower than those of the QLSTM.
\begin{table}[htbp]
\caption{\bfseries{Results: Time-Series Modeling - Bessel Function $J_2$}}
\begin{tabular}{|l|l|l|}
\hline
          & QLSTM\cite{chen2022reservoir} & QFWP \\ \hline
Epoch 1   &$1.04 \times 10^{-1}$/$1.66 \times 10^{-2}$&$1.17 \times 10^{-1}$/$1.58 \times 10^{-2}$\\ \hline
Epoch 15  &$2.30 \times 10^{-2}$/$5.35 \times 10^{-3}$&$1.22 \times 10^{-2}$/$4.56 \times 10^{-3}$\\ \hline
Epoch 30  &$1.27 \times 10^{-2}$/$2.42 \times 10^{-3}$&$5.52 \times 10^{-3}$/$7.80 \times 10^{-4}$\\ \hline
Epoch 100 &$6.97 \times 10^{-4}$/$1.21 \times 10^{-5}$&$8.57 \times 10^{-4}$/$2.30 \times 10^{-3}$\\ \hline
\end{tabular}
\label{tab:results_Bessel_compare}
\end{table}
\subsubsection{Time Series Prediction (NARMA Benchmark)}
\label{sec:time_series_prediction}
\begin{figure}[htbp]
\begin{center}
\includegraphics[width=1\columnwidth]{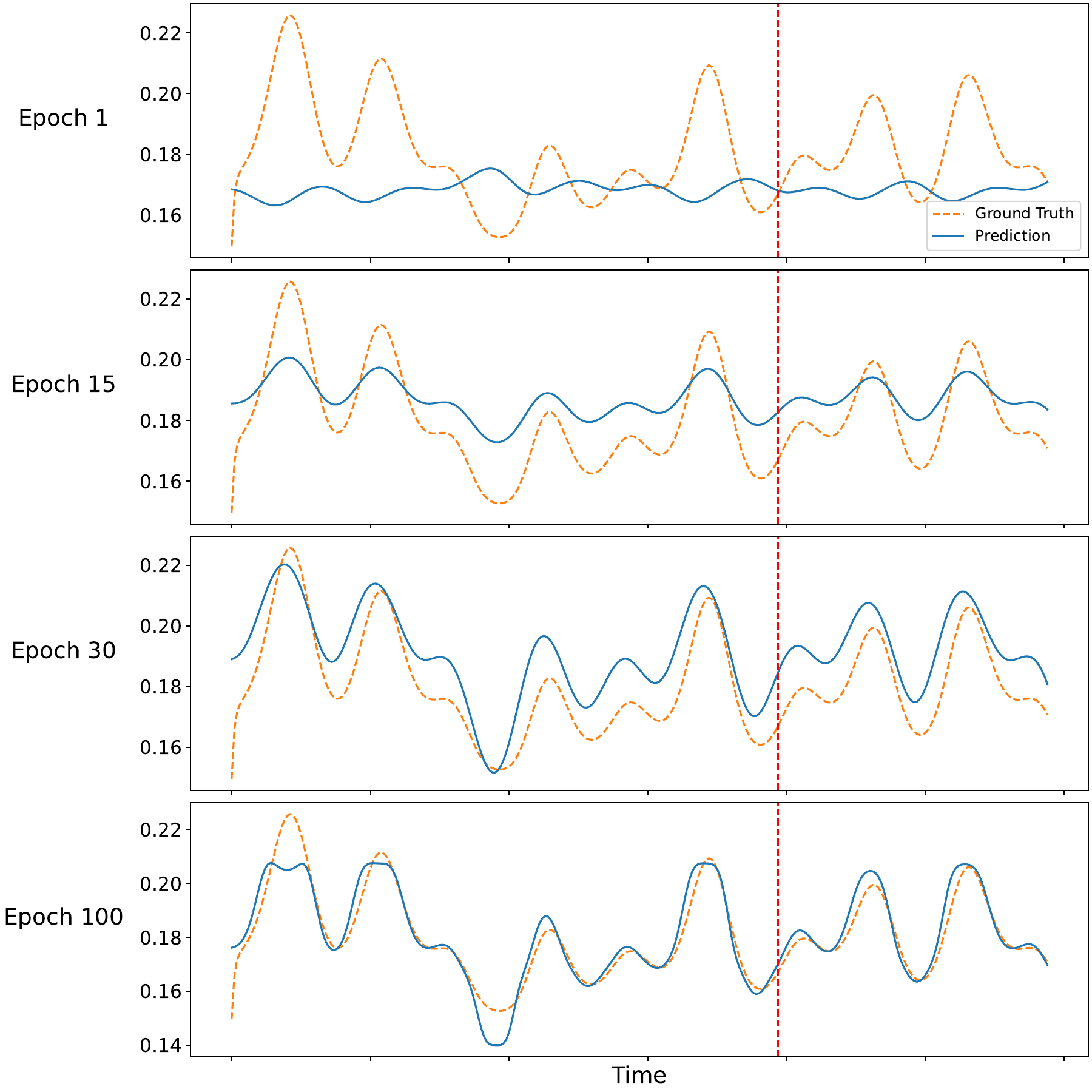}\vskip -0.1in
\caption{{\bfseries Results: Quantum FWP for NARMA5.}}
\label{fig:results_NARMA5}
\end{center}
\vskip -0.1in
\end{figure}

\begin{figure}[htbp]
\begin{center}
\includegraphics[width=1\columnwidth]{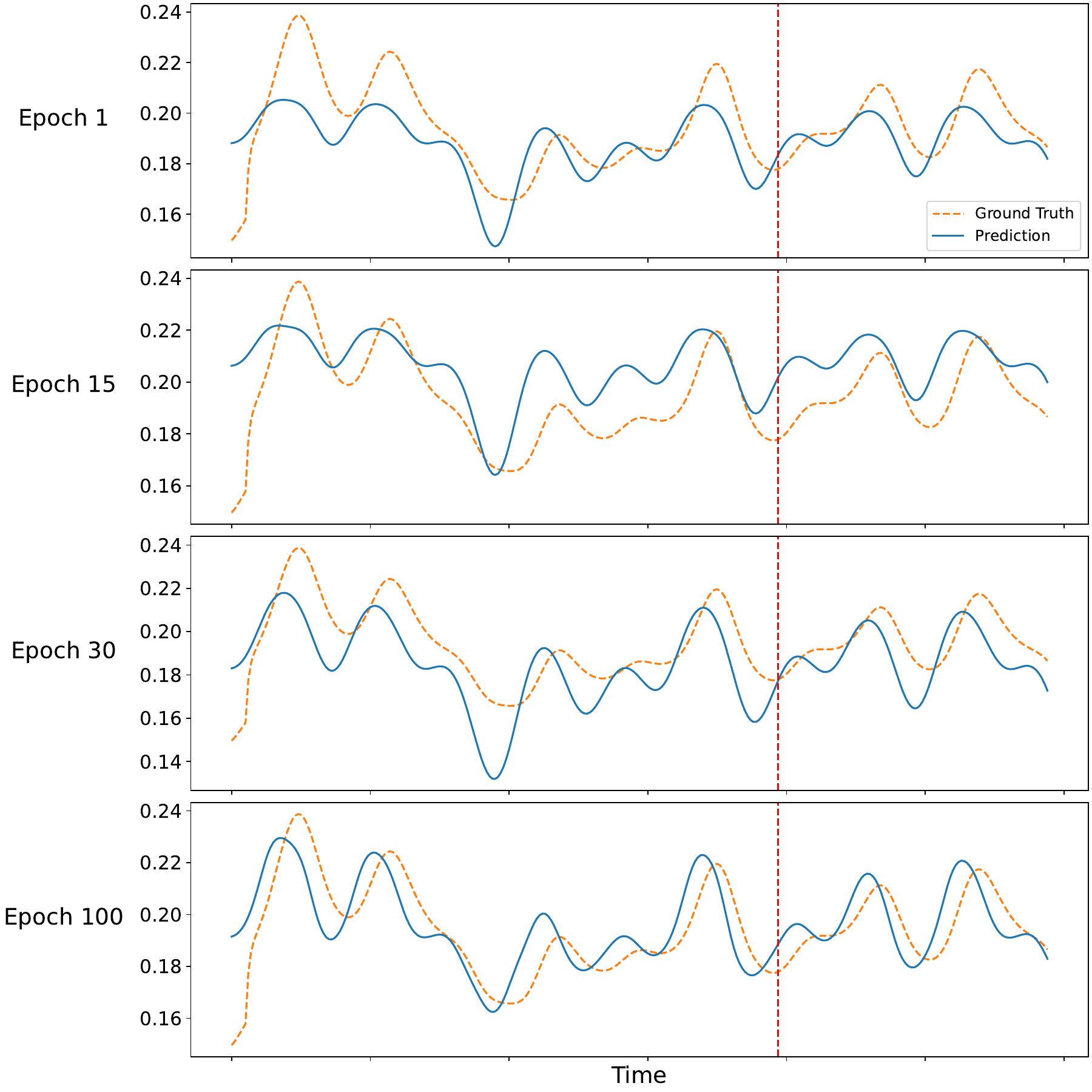}\vskip -0.1in
\caption{{\bfseries Results: Quantum FWP for NARMA10.}}
\label{fig:results_NARMA10}
\end{center}
\vskip -0.1in
\end{figure}

In this part of the simulation, we examine the NARMA (Non-linear Auto-Regressive Moving Average) time series \cite{atiya2000new} to assess the QFWP's capability in nonlinear time series modeling.
The NARMA series that we use in this work can be defined by \cite{goudarzi2014comparative}:
\begin{equation}
y_{t+1}=\alpha y_{t}+\beta y_{t}\left(\sum_{j=0}^{n_{o}-1} y_{t-j}\right)+\gamma u_{t-n_{o}+1} u_{t}+\delta
\end{equation}
where $(\alpha, \beta, \gamma, \delta)=(0.3,0.05,1.5,0.1)$ and $n_{0}$ is used to determine the nonlinearity.
The input $\left\{u_{t}\right\}_{t=1}^{M}$ for the NARMA tasks is:
\begin{equation}
u_{t}=0.1\left(\sin \left(\frac{2 \pi \bar{\alpha} t}{T}\right) \sin \left(\frac{2 \pi \bar{\beta} t}{T}\right) \sin \left(\frac{2 \pi \bar{\gamma} t}{T}\right)+1\right)
\end{equation}
where $(\bar{\alpha}, \bar{\beta}, \bar{\gamma}, T)=(2.11,3.73,4.11,100)$ as used in \cite{suzuki2022natural}. We set the length of inputs and outputs to $M = 300$. In this paper, we consider $n_{0} = 5$ and  $n_{0} = 10$, NARMA5 and NARMA10 respectively.
As demonstrated in \figureautorefname{\ref{fig:results_NARMA5}} and \figureautorefname{\ref{fig:results_NARMA10}}, the proposed QFWP successfully captures the patterns of both NARMA5 and NARMA10 after training. The complexity of the task in this scenario is notably higher than that of the previously considered Damped SHM and Bessel functions, requiring a longer time for the model to learn accurate predictions. As indicated in \tableautorefname{\ref{tab:results_NARMA5_compare}} and \tableautorefname{\ref{tab:results_NARMA10_compare}}, the QFWP model achieves performance closely aligned with the QLSTM model results, as previously reported in \cite{chen2022reservoir}. It is worth noting that the QFWP model exhibits a smaller size than QLSTM, as detailed in \tableautorefname{\ref{tab:time_series_number_params}}.
\begin{table}[htbp]
\caption{\bfseries{Results: Time-Series Modeling - NARMA5}}
\begin{tabular}{|l|l|l|}
\hline
          & QLSTM\cite{chen2022reservoir} & QFWP \\ \hline
Epoch 1   &$3.99 \times 10^{-3}$/$4.07 \times 10^{-4}$&$4.44 \times 10^{-2}$/$3.48 \times 10^{-4}$\\ \hline
Epoch 15  &$3.30 \times 10^{-4}$/$4.23 \times 10^{-4}$&$2.99 \times 10^{-4}$/$8.76 \times 10^{-5}$\\ \hline
Epoch 30  &$1.86 \times 10^{-4}$/$2.06 \times 10^{-4}$&$2.71 \times 10^{-4}$/$1.57 \times 10^{-4}$\\ \hline
Epoch 100 &$9.85 \times 10^{-5}$/$2.52 \times 10^{-5}$&$5.15 \times 10^{-5}$/$1.68 \times 10^{-5}$\\ \hline
\end{tabular}
\label{tab:results_NARMA5_compare}
\end{table}
\begin{table}[htbp]
\caption{\bfseries{Results: Time-Series Modeling - NARMA10}}
\begin{tabular}{|l|l|l|}
\hline
          & QLSTM\cite{chen2022reservoir} & QFWP \\ \hline
Epoch 1   &$4.19 \times 10^{-3}$/$4.71 \times 10^{-4}$&$4.43 \times 10^{-2}$/$9.00 \times 10^{-5}$\\ \hline
Epoch 15  &$3.35 \times 10^{-4}$/$4.73 \times 10^{-4}$&$4.95 \times 10^{-4}$/$2.24 \times 10^{-4}$\\ \hline
Epoch 30  &$3.20 \times 10^{-4}$/$3.74 \times 10^{-4}$&$2.79 \times 10^{-4}$/$1.56 \times 10^{-4}$\\ \hline
Epoch 100 &$2.59 \times 10^{-4}$/$9.50 \times 10^{-5}$&$2.58 \times 10^{-4}$/$9.70 \times 10^{-5}$\\ \hline
\end{tabular}
\label{tab:results_NARMA10_compare}
\end{table}
\subsection{Reinforcement Learning}
\subsubsection{Environments}
\begin{figure}[htbp]
\begin{center}
\includegraphics[width=1\columnwidth]{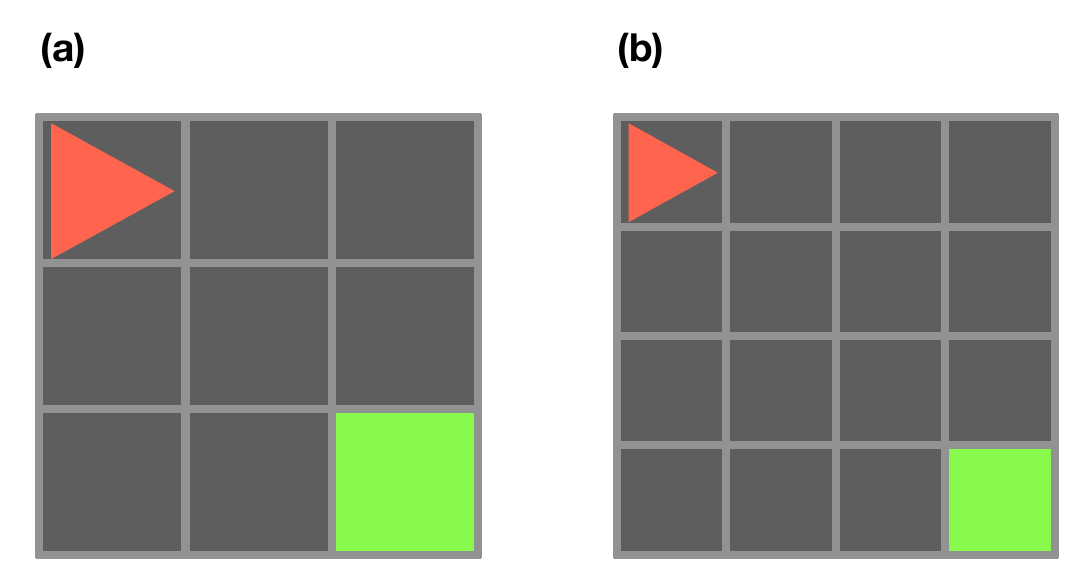}\vskip -0.1in
\caption{{\bfseries MiniGrid environments.}}
\label{fig:MiniGridEnv}
\end{center}
\vskip -0.1in
\end{figure}

Within this study, we engage the MiniGrid-Empty environment, a widely utilized maze navigation framework \cite{gym_minigrid}. The primary focus of our QRL agent is to craft effective action sequences based on real-time observations, facilitating traversal from the starting point to the green box as illustrated in \figureautorefname{\ref{fig:MiniGridEnv}}. Noteworthy is the distinctive feature of the MiniGrid-Empty environment—a $147$-dimensional observation vector denoted as $s_t$. This environment offers an action spectrum, denoted as $\mathcal{A}$, encompassing six actions: \textit{turn left}, \textit{turn right}, \textit{move forward}, \textit{pick up an object}, \textit{drop the carried object}, and \textit{toggle}. However, only the first three actions bear practical significance within this context, demanding the agent's discernment. Furthermore, successful navigation to the goal endows the agent with a score of $1$. Yet, this achievement is tempered by a penalty computed via $1 - 0.9 \times (\textit{number of steps}/\textit{max steps allowed})$, with the maximum steps permitted set at $4 \times n \times n$, predicated on the grid size $n$ \cite{gym_minigrid}. We consider the two cases with $n=5$ and $n=6$ in this study.

\subsubsection{QLSTM baseline}
\begin{figure}[htbp]
\begin{center}
\includegraphics[width=1\columnwidth]{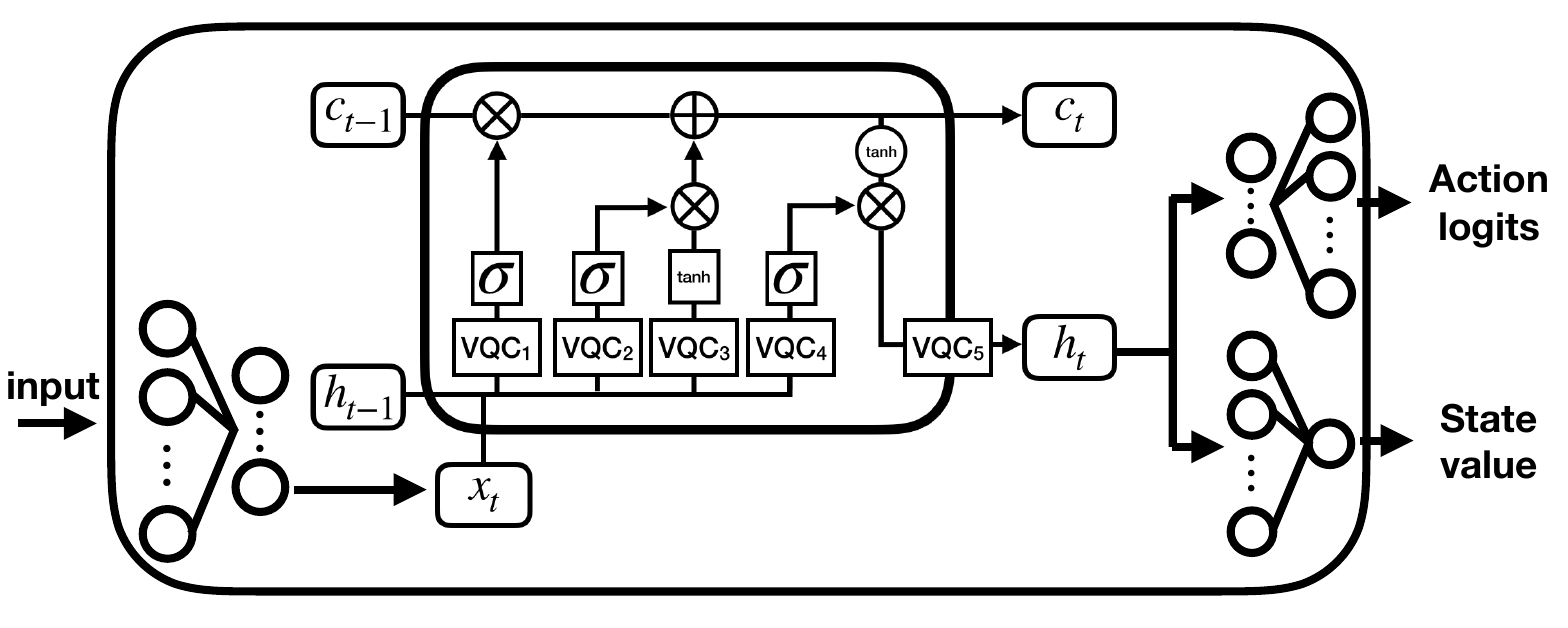}\vskip -0.1in
\caption{{\bfseries QLSTM baseline used in this work.}}
\label{fig:QLSTM_baseline}
\end{center}
\vskip -0.1in
\end{figure}

The proposed QFWP model has the capability to memorize the information of previous time-steps in the form of quantum circuit parameters. In quantum RL, quantum recurrent neural networks (QRNN) can be used to achieve the same goal. 
Here, we consider the QLSTM-based QRL agent developed in the works \cite{chen2023quantum_LSTM_RL,chen2023efficientQRL_QRC}. To improve the training efficiency, we employ the asynchronous methods \cite{mnih2016asynchronous,CHEN2023321Async}. We consider QLSTM models with different number of VQC layers to evaluate how good the proposed QFWP model is. The QLSTM models we consider in this work are of a structure as shown in \figureautorefname{\ref{fig:QLSTM_baseline}}.   
In this simulation, we employ classical neural networks (NNs) for two primary purposes: compressing the input vector to a dimension suitable for quantum simulation and processing the output from a QLSTM to derive final results. Specifically, we utilize an 8-qubit QLSTM, allocating the initial 4 qubits for input and the subsequent 4 qubits for the hidden dimension. The NN responsible for transforming the input vector into a compressed representation possesses an input dimension of 147 and an output dimension of 4 (with a number of parameters: $147 \times 4 + 4 = 592$). Additionally, we incorporate two other NNs into the system: one for translating QLSTM outputs into action logits (referred to as the "actor") and another for determining the state value (referred to as the "critic"). Both NNs share an input dimension equivalent to the hidden dimension of the QLSTM (which in this case is 4). 
The actor NN has an output dimension of 6, while the critic NN has an output dimension of 1. The number of parameters for the actor NN is calculated as $4 \times 6 + 6 = 30$, and for the critic NN, it is $4 \times 1 + 1 = 5$. The number of quantum circuit parameters in QLSTM with $n$ VQC layer is: $8 \times 3 \times 5 \times n = 120n$ in which the VQCs are 8-qubit and each general rotation gate is parameterized by 3 parameters (angles). There are $5$ VQCs in a QLSTM.
The detailed description of QLSTM can be found in \cite{chen2022quantumLSTM,chen2022reservoir}. We summarize the number of parameters of QLSTM we compare in this work in \tableautorefname{\ref{tab:rl_number_params}}.

\subsubsection{QFWP in RL}
\begin{figure}[htbp]
\begin{center}
\includegraphics[width=1\columnwidth]{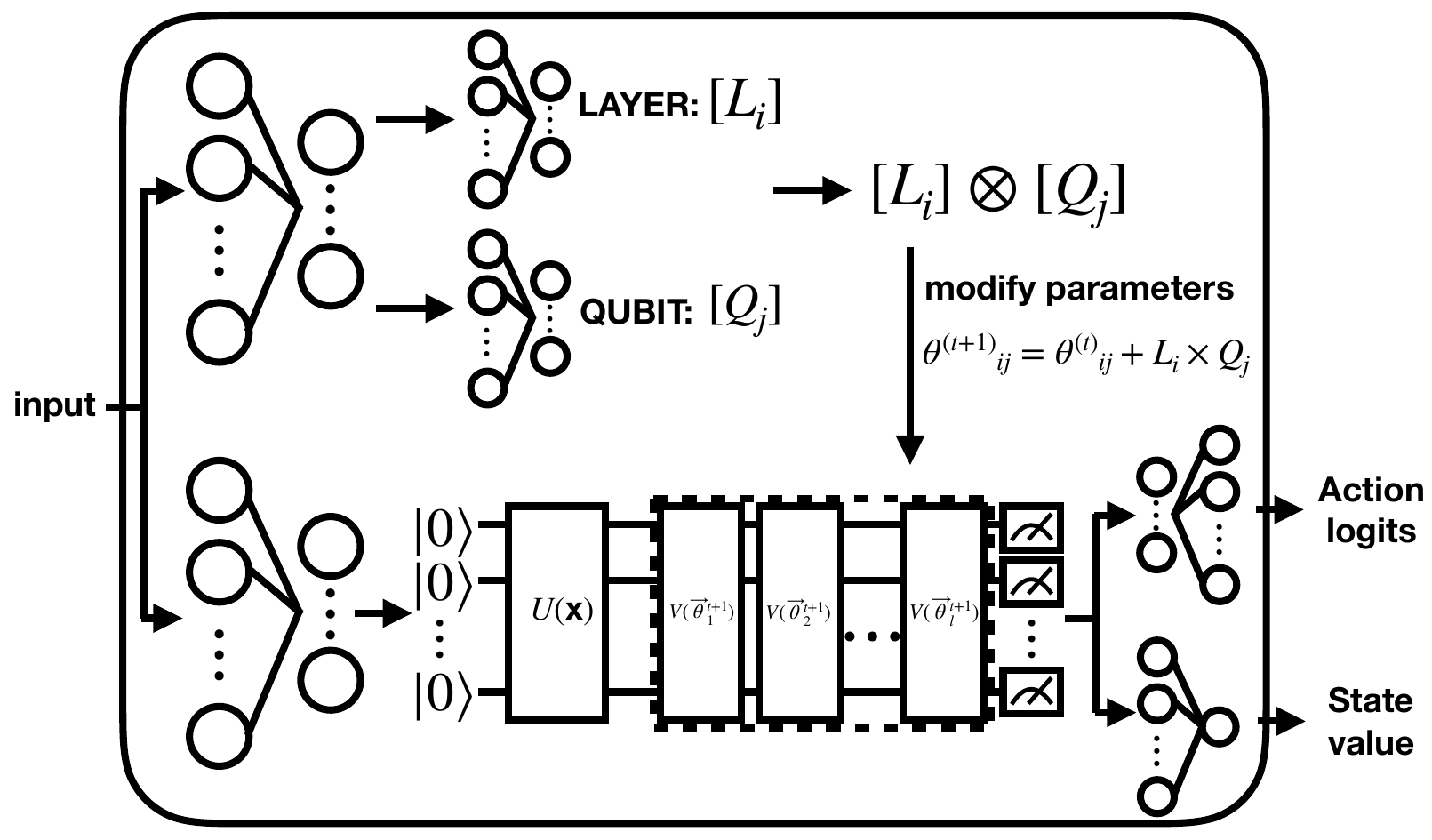}\vskip -0.1in
\caption{{\bfseries QFWP as a RL agent.}}
\label{fig:QFWP_as_RL_agent}
\end{center}
\vskip -0.1in
\end{figure}

The QFWP used in this part of the simulation includes a classical NN-based  \emph{slow programmer} which is consisted of a \emph{slow programmer encoder} (with number of parameter $ (\textbf{state dim} + 1) \times  \textbf{latent dim}$) and two separate NN for generating parameter updates for different quantum layer (with number of parameter $ (\textbf{latent dim}+1)  \times \textbf{num VQC layer}$) and qubit indexes (with number of parameter $ (\textbf{latent dim}+1) \times \textbf{num qubits}$) as shown in \figureautorefname{\ref{fig:vqFWP_Concept}}. The $\textbf{state dim}$ and $\textbf{num qubits}$ here are set to be $8$. 
There are $147 \times 8 + 8 = 1184$ parameters in the slow programmer encoder, $(8+1) \times \textbf{num VQC layer}$ in the NN for different quantum layers and $8 \times 8 + 8 = 72$ in the NN for qubit indexes. 
The \emph{fast programmer} is consisted of a VQC with architecture described in \figureautorefname{\ref{fig:detailed_VQC}}. We consider $8$-qubit VQCs with $L=2$ and $L=4$ in this part of simulation. The number of quantum parameters are therefore $16$ and $32$, respectively. In addition, there is a NN for transforming input observation vector into a compressed vector suitable for existing VQC simulation. The number of parameters for this NN is $147 \times 8 + 8 = 1184$.  
Outside the main QFWP model, there are two NN for processing the outputs from QFWP to generate the final action logits and state values. The number of parameters of these two NNs are $8 \times 6 + 6 = 54$ and $8 \times 1 + 1= 9$, respectively.
The entire architecture of the QFWP as a RL agent is depicted in \figureautorefname{\ref{fig:QFWP_as_RL_agent}}.
We summarize the number of parameters of QFWP in \tableautorefname{\ref{tab:rl_number_params}}.
\begin{table}[htbp]
\caption{\bfseries{Number of parameters in QFWP and QLSTM models in QRL experiments.}}
\begin{center}
    
\begin{tabular}{|l|l|l|}
\hline
                        & Classical & Quantum \\ \hline
QLSTM-2 VQC Layer       & 627       & 240     \\ \hline
QLSTM-4 VQC Layer       & 627       & 480     \\ \hline
QLSTM-6 VQC Layer       & 627       & 720     \\ \hline
QLSTM-8 VQC Layer       & 627       & 960     \\ \hline
QLSTM-10 VQC Layer      & 627       & 1200    \\ \hline
Quantum FWP-2 VQC Layer & 2521      & 16      \\ \hline
Quantum FWP-4 VQC Layer & 2539      & 32      \\ \hline
\end{tabular}
\end{center}
\label{tab:rl_number_params}
\end{table}

\subsubsection{Hyperparameters}
The hyperparameters for the proposed QFWP in RL with QA3C training \cite{CHEN2023321Async,chen2023efficientQRL_QRC} are configured as follows: Adam optimizer with a learning rate of $1 \times 10^{-4}$, $beta_{1} = 0.92$, $beta_{2} = 0.999$, model lookup steps denoted as $L = 5$, and a discount factor $\gamma = 0.9$. In the QA3C training process, the local agents or models calculate their own gradients every $L$ steps, which corresponds to the length of the trajectory used during model updates. 
The number of parallel processes (local agents) is $80$.
We present the average score along with its standard deviation over the past 5,000 episodes to illustrate both the trend and stability.
\begin{figure}[htbp]
\begin{center}
\includegraphics[width=1\columnwidth]{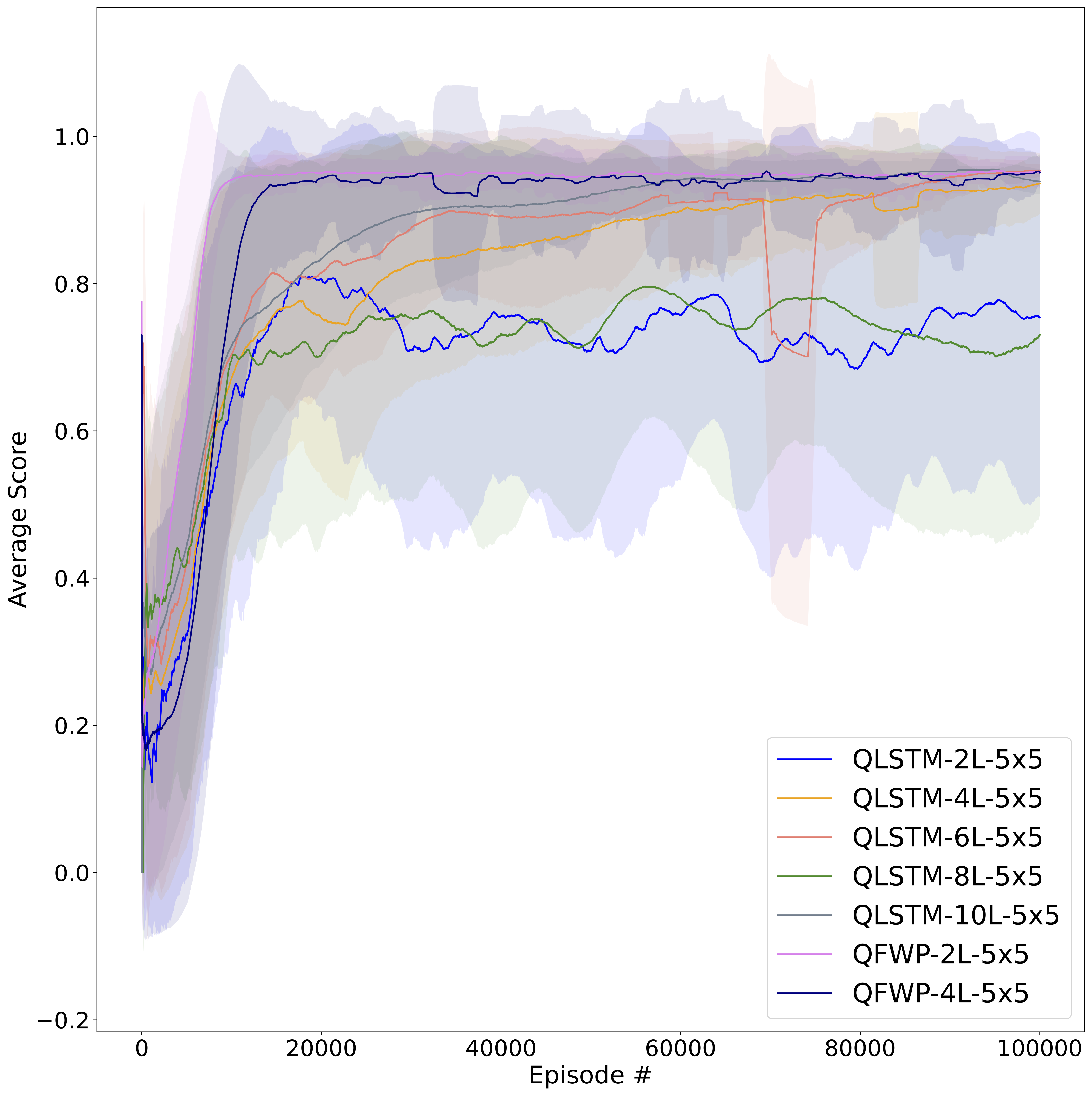}\vskip -0.1in
\caption{{\bfseries Results: Quantum FWP in MiniGrid-Empty-5x5 environment.}}
\label{fig:results_5x5}
\end{center}
\vskip -0.1in
\end{figure}
\subsubsection{Results}
\textbf{MiniGrid-Empty-5x5} As depicted in \figureautorefname{\ref{fig:results_5x5}}, the QFWP with two and four VQC layers surpasses all considered QLSTM models. The QFWP not only attains higher scores but also achieves these results more rapidly, as measured by the number of training episodes. For instance, the QLSTM model with 10 VQC layers eventually achieves the optimal score, but this accomplishment requires approximately 60,000 episodes. Additionally, we observe that QFWP models, once reaching optimal scores, exhibit greater stability compared to QLSTM models.\\
\begin{figure}[htbp]
\begin{center}
\includegraphics[width=1\columnwidth]{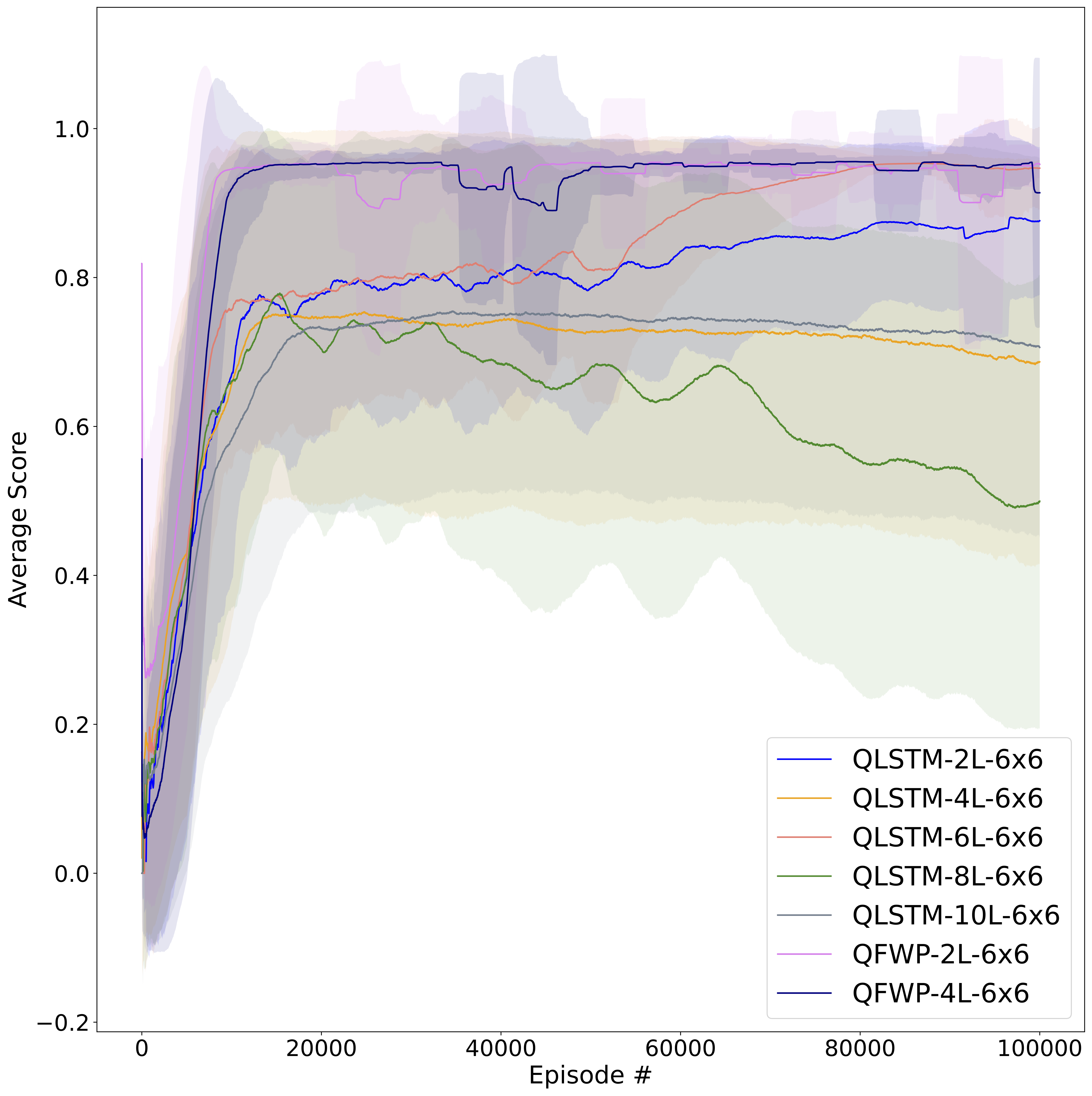}\vskip -0.1in
\caption{{\bfseries Results: Quantum FWP in MiniGrid-Empty-6x6 environment.}}
\label{fig:results_6x6}
\end{center}
\vskip -0.1in
\end{figure}
\textbf{MiniGrid-Empty-6x6} As illustrated in \figureautorefname{\ref{fig:results_6x6}}, the QFWP with two and four VQC layers outperforms all considered QLSTM models. The QFWP not only attains higher scores but also achieves these results more rapidly, as indicated by the number of training episodes. For instance, the QLSTM model with 6 VQC layers also reaches the optimal score, but this accomplishment requires approximately 80,000 episodes. Furthermore, we observe that QFWP models, once achieving optimal scores, exhibit greater stability compared to QLSTM models.
\section{\label{sec:Conclusion}Conclusion}
In this study, we introduce a hybrid quantum-classical framework of fast weight programmers for time-series modeling and reinforcement learning. Specifically, classical neural networks function as slow programmers, generating updates or changes for fast programmers implemented through variational quantum circuits. We assess the proposed model in time-series prediction and reinforcement learning tasks. Our numerical simulation results demonstrate that our framework achieves time-series prediction capabilities comparable to fully trained quantum long short-term memory (QLSTM) models reported in prior works. Furthermore, our model exhibits superior performance in navigation tasks, surpassing QLSTM models with higher stability and average scores under quantum A3C training. The proposed approach establishes an efficient avenue for pursuing hybrid quantum-classical sequential learning without the need for quantum recurrent neural networks.

\bibliographystyle{IEEEtran}
\bibliography{apssamp,bib/fwp,bib/qml_examples,bib/tool,bib/narma,bib/qrl,bib/metalearning,bib/vqc,bib/qc_basic,bib/classical_ml}

\end{document}